\documentclass{elsart}

\usepackage{graphicx}  
\usepackage{dcolumn}   
\usepackage{bm}        
\usepackage[english]{babel}
\usepackage{amsfonts,amsmath,amssymb,mathrsfs}
\usepackage{epstopdf}
\usepackage{subfigure}
\usepackage{color}

\def\a{\alpha}
\def\r{\rho}
\def\s{\sigma}
\def\t{\tau}
\def\m{\mu}
\def\n{\nu}
\def\k{\kappa}
\def\th{\theta}
\def\g{\gamma}\def\G{\Gamma}
\def\L{\Lambda}\def\l{\lambda}
\def\D{\Delta}
\def\la{\langle}
\def\ra{\rangle}
\def\o{\omega}\def\O{\Omega}
\def\d{\delta}
\def\p{\partial}

\def\z{\zeta}
\def\de{\partial}
\def\nn{\nonumber}

\def\half{\textstyle{\frac{1}{2}}}

\def\bdoc{\begin{document}}
\def\edoc{\end{document}}

\def\beq{\begin{equation}}
\def\eeq{\end{equation}}
\def\bea{\begin{eqnarray}}
\def\eea{\end{eqnarray}}
\def\ben{\begin{enumerate}}
\def\een{\end{enumerate}}
\def\la{\langle}
\def\ra{\rangle}
\def\a{\alpha}
\def\b{\beta}
\def\g{\gamma}\def\G{\Gamma}
\def\d{\delta}\def\D{\Delta}
\def\e{\epsilon}

\def\th{\theta}
\def\k{\kappa}
\def\l{\lambda}
\def\m{\mu}
\def\n{\nu}
\def\o{\omega}
\def\p{\pi}
\def\r{\rho}
\def\s{\sigma}
\def\t{\tau}
\def\L{{\mathcal L}}
\def\S{\Sigma }
\def\gsim{\; \raisebox{-.8ex}{$\stackrel{\textstyle >}{\sim}$}\;}
\def\lsim{\; \raisebox{-.8ex}{$\stackrel{\textstyle <}{\sim}$}\;}
\def\gtrsim{\gsim}
\def\lessim{\lsim}
\def\loc{{\rm local}}
\def\vm{v_{\rm max}}
\def\bh{\bar{h}}
\def\del{\partial}
\def\nab{\nabla}
\def\half{{\textstyle{\frac{1}{2}}}}
\def\fourth{{\textstyle{\frac{1}{4}}}}

\def\bD{{\bf D}}
\def\bE{{\bf E}}
\def\bF{{\bf F}}
\def\bB{{\bf B}}
\def\bP{{\bf P}}
\def\bV{{\bf v}}
\def\bv{{\bf v}}
\def\bx{{\bf x}}
\def\by{{\bf y}}
\def\bz{{\bf z}}
\def\ba{{\bf a}}
\def\bd{{\bf d}}
\def\bs{{\bf s}}
\def\bn{{\bf n}}
\def\bp{{\bf p}}

\def\O{\Omega}

\def\br{{\bf r}}
\def\bnab{{\bf \nab}}

\def\tE{\tilde{E}}
\def\tL{\tilde{L}}

\newcommand{\scri}{\mathscr{I}}
\newcommand{\sun}{\ensuremath{\odot}}%
\def\e{{\mathrm e}}%
\def\g{{\mbox{\sl g}}}%
\def\Box{\nabla^2}%
\def\d{{\mathrm d}}%
\def\R{{\rm I\!R}}%
\def\ie{{\em i.e.\/}}%
\def\eg{{\em e.g.\/}}%
\def\etc{{\em etc.\/}}%
\def\etal{{\em et al.\/}}%
\def\HRULE{{\bigskip\hrule\bigskip}}

\def\d{{\mathrm{d}}}
\def\J{{\mathscr{J}}}
\def\L{{\mathscr{L}}}
\def\H{{\mathscr{H}}}
\def\T{{\mathscr{T}}}
\def\V{{\mathscr{V}}}

\def\sech{{\mathrm{sech}}}

\begin{document}
\begin{frontmatter}


\title{Strong coupling in extended Ho\v rava--Lifshitz gravity}
\author{Antonios Papazoglou$^{1}$ and Thomas P. Sotiriou$^{2}$}
\address{$^{1}$Institute of Cosmology and Gravitation, University of Portsmouth, Portsmouth, PO1 3FX, UK\\
$^{2}$Department of Applied Mathematics and Theoretical Physics, Centre for  
Mathematical Sciences, University of Cambridge, Wilberforce Road,  
Cambridge, CB3 0WA, UK}
\begin{abstract}
An extension of Ho\v rava--Lifshitz gravity was recently proposed in order to address the pathological behaviour of the scalar mode all previous versions of the theory exhibit. We show that even in this new extension the strong coupling persists, casting doubts on whether such a model can constitute an interesting alternative to general relativity (GR).
\end{abstract}  
\begin{keyword}
Ho\v rava--Lifshitz gravity \sep strong coupling \sep scalar mode \sep Lorentz violations
\PACS 04.60.-m \sep 04.50.Kd \sep  11.25.Db \sep 11.30.Cp
\end{keyword}
\end{frontmatter}

A new perspective towards addressing the quantum gravity puzzle has received increased attention recently: Ho\v rava proposed a theory of gravity which does not respect Lorentz invariance \cite{Horava:2009uw}. Instead, it assumes the existence of a preferred foliation by 3-dimensional constant time hypersurfaces, which splits spacetime into space and time. This allows to add higher order spatial derivatives of the metric to the action, without introducing higher order time derivatives. This is supposed to improve the ultraviolet (UV) behavior of the graviton propagator and render the theory power-counting renormalizable without introducing ghost modes, which are common when adding higher order curvature invariants to the action in a covariant manner \cite{Stelle:1977ry}. The new theory is supposed to be an adequate UV completion of GR, above its natural cutoff scale, which is of the  order of the Planck scale.

Such a theory cannot be invariant under the full set of diffeomorphisms, but it can still be invariant under the more limiting foliation-preserving diffeomorphisms, $t\to\tilde{t}(t),\;x^i\to\tilde{x}^i(t,x^i)$. In this setting, it is natural to consider the Arnowitt--Deser--Misner (ADM) decomposition of spacetime
\begin{equation}
\d s^2 = - N^2 c^2 \d t^2 + g_{ij}(\d x^i + N^i \d t) (\d x^j + N^j \d t).
\end{equation}
Defining the extrinsic curvature as
\begin{equation}
K_{ij} = {1\over2N} \left\{  \dot g_{ij} - \nabla_i N_j - \nabla_j N_i \right\},
\end{equation}
the action of the theory is of the form
\begin{equation}
S=\frac{M_{\rm pl}^2}{2}\int \d^3x \d t N \sqrt{g} \left\{ K^{ij} K_{ij} - \lambda K^2 -\V\right\}\, ,
\end{equation}
where $M_{\rm pl}$ is the Planck mass, $g$ is the determinant of the spatial metric $g_{ij}$ and $\lambda$ is a dimensionless running coupling. $\V$ depends only on $g_{ij}$ and its spatial derivatives.

Power-counting renormalizability requires that $\V$ include at least sixth order terms in derivatives of $g_{ij}$. To see this, it is better to switch from the $c=1$ units used above, to units that impose the scaling
\beq
[\d t]=[\kappa]^{-3}, \quad [\d x]=[\kappa]^{-1},
\eeq
where $\kappa$ is a placeholder symbol with dimensions of momentum. This makes the couplings of the kinetic term and the sixth order derivatives dimensionless and renders the theory power counting renormalizable.\footnote{See also \cite{Visser:2009fg} for a more extensive discussion of this principle, using the simpler case of a scalar field as an example.} Switching back to $c=1$ units, which are more suitable for discussing the infrared behavior, $\V$ has the general structure
\beq
\V=2\Lambda - R+ M_{\rm pl}^{-2}O(R^2)+M_{\rm pl}^{-4} O(R^3), \label{pot} 
\eeq
where, $R$ is the Ricci scalar of $g_{ij}$ and $O(R^2)$ and $O(R^3)$ denote all terms that can be constructed from $g_{ij}$ and its spatial derivatives up to dimensions 4 and 6 respectively. $\Lambda$ plays the role of a cosmological constant and can be easily expressed in terms of a dimensionless coupling as $\Lambda=g_0 M_{\rm pl}^2/2$, whereas the coefficient of $R$ has been set to $-1$ by a suitable rescaling of the coordinates. Note that the {\em natural} scale suppressing higher order operators here is the Planck scale  $M_{\rm pl}$; we will return to this point later. From a phenomenological perspective, the hope is that $\lambda$ will flow to $1$ in the IR, which is the general relativistic  value, and considering that the higher order terms will be suppressed, Lorentz invariance will be restored as an emergent symmetry.

In principle $\V$ should include all possible terms up to a suitable dimension. Ho\v rava proposed  a specific structure for $\V$ inspired by condensed matter systems, dubbed ``detailed balance'', in order to reduce the number of terms \cite{Horava:2009uw}. Since its physical motivation is unclear and it leads to the wrong sign of the cosmological constant  (see however \cite{pope}), we will not restrict ourselves to it here. In any case, our results will be independent of the specific choice of $\V$. 

Another restricted version of the theory is the one based on the assumption of ``projectability'', $N=N(t)$. This was proposed as a way to match the reduced symmetry and still be able to set $N\to 1$ using gauge transformations. This assumption significantly reduces the number of terms in $\V$. It has been used in Refs.~\cite{Sotiriou:2009gy,Sotiriou:2009bx} as a simplicity assumption in order to construct and study the most general action in this framework.\footnote{See also \cite{Kiritsis:2009sh} for a more cosmologically oriented study.}

As already pointed out in Ref.~\cite{Horava:2009uw},  the breaking of general covariance generically introduces a scalar degree of freedom. The behavior of this scalar creates a serious viability issue for all versions of the theory.  To be specific, in the version with ``projectability'' and for maximally symmetric backgrounds, the scalar is (classically) unstable for  $\lambda>1$ or $\lambda<1/3$ and it becomes a ghost (quantum-mechanically unstable)
for $1/3<\lambda<1$  \cite{Sotiriou:2009bx,Koyama:2009hc,Bogdanos:2009uj}. 
For the same backgrounds, when $\lambda \to 1$   the cubic interactions blow up and the scalar becomes strongly coupled \cite{Blas:2009yd,Koyama:2009hc}.

The non-projectable version is also burdened with similar problems. Even though the scalar mode appears to be absent (frozen) from the spectrum for backgrounds that are either time-independent or spatially homogeneous ({\it e.g.} maximally symmetric or cosmological  backgrounds \cite{Gao:2009ht,Blas:2009yd}),   it reveals itself for more complicated backgrounds \cite{Blas:2009yd} and signals the breakdown of the theory at unacceptably long distances, leading to strong coupling and instabilities. See also  \cite{Charmousis:2009tc,Li:2009bg} regarding pathologies of the non-projectable case.

Having in mind that the problems in the non-projectable case stem from the anomalous structure of the quadratic action for the scalar degree of freedom around smooth backgrounds, Blas, Pujolas and Sibiryakov recently proposed an extension of the theory which leads to a healthy quadratic action  \cite{Blas:2009qj}.  Our purpose here is to examine whether this extended version of Ho\v rava-Lifshitz gravity also avoids the strong coupling issues burdening its predecessors.

We start by reviewing the extended theory proposed in \cite{Blas:2009qj}.
The key idea is to add  to  $\V$ terms with up to six spatial derivatives, constructed with the 3-vector
\beq
a_i\equiv \partial_i N/N,
\eeq
and its contractions with 3-dimensional curvature invariants of $g_{ij}$. Such terms are manifestly invariant under the symmetry of the initial theory, $t\to\tilde{t}(t),\;x^i\to\tilde{x}^i(t,x^i)$. In principle, one can straightforwardly write down all of the new operators  of dimensions 4 and 6 that involve $a_i$ and its contractions with 3-curvature invariants and modify the theory in the UV. These terms are numerous and we will avoid listing them here. However, note that there are more terms\footnote{Since $N=N(t,x^i)$, integration by parts does not allow to drop as many terms as one would expect. For instance, $\nabla^2 R$ is no longer a surface term.} than those contributing to the quadratic action listed in Ref.~\cite{Blas:2009qj}.

In the following, we will focus on the IR limit of the theory. This means that we will take into account operators in $\V$ of lowest order, \ie ~of at most dimension 2.  Apart from the first two terms  in (\ref{pot}),  there is only one more term involving $a^i$ that has to be added,
\beq
\eta~ a_i a^i , \label{newcoupling}
\eeq
where $\eta$ is a coupling constant, which is in principle also running, just as $\lambda$. In fact, the full IR action, in which we will focus from now on, is
(setting $M_{\rm pl}=1$)
\beq
\label{IRaction}
\! S=\frac{1}{2}\int \d^3x \d t  N \sqrt{g} \left\{ K^{ij} K_{ij} - \lambda K^2 +R + \eta \, a_i a^i  \right\}.
\eeq

To study the interactions of the theory, we will adopt a gauge similar to the one used in Ref.~\cite{Koyama:2009hc}  and write down the scalar perturbations of the metric as
\beq
N=e^{\a(t,x)}, \  \  \  N_i=\de_i \b(t,x),  \  \  \   g_{ij}=e^{2\z(t,x)} \delta_{ij} . \label{gauge}
\eeq
The above form differs from the most general scalar perturbation possible, by  the perturbation of $g_{ij}$ of the form
$2 \de_i \de_j E(t,x)$.
Such a  perturbation transforms under the foliated diffeomorphism group as $E \to E - \chi$ \cite{Wang:2009yz}, where  $x^i \to x^i + \chi^{,i}$.  It is then straightforward  to use the freedom in $\chi(t,x)$  in order to set $E=0$. 

Substituting eqs.~(\ref{gauge}) into eq.~(\ref{IRaction}) leads to 
the quadratic Lagrangian, which after partial integrations reads
\beq
{\mathcal L}_2= {3 \over 2}(1-3\l)\dot{\z}^2 + (\de \z)^2 +{1 \over 2} (1-\l)(\D \b)^2 - (1-3\l)\dot{\z} \D \b -2\a \D \z -{\eta \over 2}\a \D \a \label{quadL}.
\eeq
This is the quadratic Lagrangian of Ref.~\cite{Koyama:2009hc}, where the projectable version of Ho\v rava--Lifshitz gravity was considered, with two important differences. Since $N$ is now spacetime dependent,  $\a$ is  spacetime dependent as well and, therefore, the penultimate term in eq.~(\ref{quadL}) is not a total derivative. Secondly, the extended theory has the extra coupling (\ref{newcoupling}) which contributes to the quadratic action.

Varying the quadratic action with respect to $\b$ and $\a$, we obtain the momentum and the Hamiltonian constraints (assuming regular boundary conditions) 
\beq
\label{beta}
\D \b = -{1 \over c_\z^2}\dot{\z}, \qquad
\a=-{2 \over \eta} \z. 
\eeq
In the above, we used the quantity 
\beq
c_\z^2={1-\lambda \over 3\lambda -1},
\eeq
whose physical significance we will see shortly.

Substituting these constraints to the quadratic action and with appropriate partial integrations, we obtain the following action for the physical mode $\z$
\beq
S_2=- \int  \d^3x \d t  \left[{1 \over c_\z^2}\dot{\z}^2 - {\eta- 2 \over \eta }(\de \z)^2  \right ] . \label{quad}
\eeq

This is in agreement with Ref.~\cite{Blas:2009qj}, if we ignore the higher order operators (of dimension 4 and 6) which were  kept there. In the original theory, without the $\eta$ operator, it is evident from (\ref{quad}) that $c_\z$ would be the speed of sound for the perturbation $\z$. Also, from the same action, we see that without  the $\eta$ contribution,  $\zeta$ would either be unstable when $c_\z^2<0$, or it would be a ghost when $c_\z^2>0$, as mentioned above. The improved theory, however,   can cure this problem for 
\beq 
\label{constr}
c_\z^2<0 \ \ \ \ {\rm and} \ \ \ \ 0<\eta<2.
\eeq

We now turn our attention to the next order in perturbation analysis in order to check whether strong coupling can be avoided.
The cubic Lagrangian reads
\bea
{\mathcal L}_3&=&{9 \over 2}(1-3\l)\z\dot{\z}^2 -\z (\de \z)^2 -\z^2 \D \z  -(1-3\l)\z \dot{\z} \D \b - (1-3\l) \dot{\z}\de_k \z \de^k \b \nn \\
&&+(1-\l) \D \b \de_k \z \de^k \b  -\frac{1}{2} \z \de_i \de_j \b \de^i \de^j \b - 2 \de_i \de_j \b \de^i \b \de^j \z  + {\l \over 2} \z (\D \b)^2 \nn \\
&&-{3 \over 2}(1-3\l)\a \dot{\z}^2+ (1-3\l)\a \dot{\z}\D \b -\frac{1}{2}\a \de_i \de_j \b \de^i \de^j \b +{\l \over 2}\a (\D \b)^2 -\a (\de \z)^2 \nn \\
&&-2 \a \z \D \z -\a^2 \D \z  + {\eta \over 2} \z (\de \a)^2 + {\eta \over 2} \a (\de \a)^2,
\eea

Substituting the constraints (\ref{beta}), 
in the same fashion as above for the quadratic action, leads to the cubic action 
\bea
\!\!\!S_3\!&=&\!\int \d t \d^3x \left\{\left( 1-\frac{4(1-\eta)}{ \eta^2}  \right) \z (\de \z)^2  - {2 \over c_\z^4} \dot{\z} \de_i \z {\de^i \over \D} \dot{\z}     \right. \nn \\
&& \left.~~~~~~~~~~~~~~~~~~~ +  \left(\frac{3}{2}+ {1 \over \eta} \right)\! \left[{1 \over  c_\z^4}    \z \left( {\de_i \de_j \over  \D}\dot{\z}\right)^2 \!- {(2 c_\z^2+1)  \over  c_\z^4}  \z \dot{\z}^2   \right]
  \right\} \label{cubic}.
  \eea
 We note again that  when if $\eta\to \infty$, we obtain the couplings found in Ref.~\cite{Koyama:2009hc} (once the momentum constraint is substituted) for the projectable case.\footnote{The projectable theory limit as noted in Ref.~\cite{Blas:2009qj} is obtained for $\eta \to \infty$. }

After restoring $M_{\rm pl}$ and performing canonical normalization of the kinetic term in the quadratic action, eq.~(\ref{quad}), as $\z= |c_\z| \hat{\z}/M_{\rm pl}$, all the terms but the first in the cubic action, eq.~(\ref{cubic}), scale as $(|c_\z| M_{\rm pl})^{-1}$. Therefore there is strong coupling for $c_\z \to 0$ ($\l \to 1$), {\it i.e.} the $\z$-interactions become strong for energies above the scale $|c_\z| M_{\rm pl}$. We note here that all terms that blow up in that limit have time derivatives of $\z$.

In order to see if the  presence of the new coupling $\eta$ can cure the strong coupling problem found above, 
we should examine whether its contributions could in general be used to  cancel (with fine-tuning, since $\eta$ is in principle running) the troublesome interactions. This can happen for the interactions in (\ref{cubic}) that depend on 
$\eta$. However, in these cases, the values of $\eta$ required for the cancelation to take place lie outside the region $0< \eta <2$, contrary to what is needed in order to have a healthy mode. Moreover, even if such a cancelation does happen, no value of $\eta$ can erase {\it all} divergent terms in the limit $c_\z \to 0$.  The third term in  (\ref{cubic}) in particular is $\eta$-independent and cannot be canceled for any value of $\eta$. Therefore, the terms that blow up at cubic order cannot  be removed from the action of the theory, not even with fine tuning of the new coupling $\eta$.  
 
The higher order terms of dimension 4 and 6 which we have ignored, can also not  be used to cancel out the troublesome interactions. This is because  they generate  new couplings involving only spatial derivatives of $\z$ and $\a$, but {\it not} $\b$.  Only the latter is related to time derivatives of $\z$ whose interactions exhibit strong coupling. An effect the higher-dimensional operators could have is the following: if instead of assuming they are suppressed by powers of the $M_{\rm pl}$ (which is the natural scale here)  we introduce a new scale $Z$ exactly for this purpose --- which is equivalent to tuning the dimensionless couplings --- then, if $Z<|c_\z| M_{\rm pl}$, the perturbative expansion will be dominated by the higher-dimensional operators before strong coupling kicks in.\footnote{This possibility was brought to our attention by the authors of Ref. \cite{Blas:2009qj} in private communication and was subsequently the subject of Ref.~\cite{Blas:2009ck}.} We will discuss this possibility, which clearly requires tuning, separately in what follows.

So far we have shown that there is inevitable strong coupling as $\lambda\to 1$, at least as long as a new scale $Z$ is not introduced. It now remains to discuss how physically relevant is this strong coupling region of the theory. First of all, just by inspecting the IR action (\ref{IRaction}), it is straightforward to see, that Lorentz invariance and diffeomorphism invariance are recovered when $\lambda\to 1$ and $\eta\to 0$. Indeed, in the initial version of the non-projectable  Ho\v rava--Lifshitz gravity where $\eta=0$, it was hoped that $\lambda$ would flow to $1$ in the infrared in order for Lorentz invariance to emerge. Clearly, this is not an option here, even if $\eta$ would flow to $0$ for some reason, as the resulting theory would be strongly coupled  at all scales.

The only other alternative is to have a theory which exhibits Lorentz violations all the way to the IR regime, as $\lambda$ does not flow to $1$ there (and neither does $\eta$ flow to $0$). In general, this is probably more likely anyway, as there is no reason to expect that the renormalization group flow would work to the benefit of Lorentz invariance recovery. However, in that case, how far from $\lambda=1$ and $\eta=0$ are we allowed to be, from an experimental perspective? Gravitational experiments, and also non-gravitational ones once matter is consistently added to the theory, should be able to provide tight constraints. This is an avenue worth exploring further.

However, for the time being it suffices to consider for purely demonstrative purposes a specific constraint already examined in  \cite{Kehagias:2009is,Blas:2009qj}: the one coming from the discrepancy between the effective gravitational constant as measured in a gravitational experiment $G_{\rm eff}$ and that measured in cosmology, $G_{\rm cosmo}$. As pointed out in  \cite{Blas:2009qj}, when considering a static metric around a point source one can start from the quadratic Lagrangian (\ref{quadL}) supplemented by the source term 
${\mathcal L}_m = T^{00} \delta N= -m \a \delta^{(3)}({\bf x})$,
and straightforwardly derive the Poisson equation (after reworking the Hamiltonian constraint in the presence of the source coupling)
\beq
\Delta \zeta=-\frac{m \delta^3(\bold{x})}{2 (1-\eta/2)}.
\eeq
Simple comparison with the usual form of the Poisson equation reveals that
\beq
G_{\rm eff}=\frac{1}{8\pi (1-\eta/2)}.
\eeq
On the other hand, whenever $N$ is not a function of time, $a^i$ vanishes, and this is the case in cosmology, where one assumes that the universe is  homogeneous and isotropic at large scales. Thus, the cosmological equations will be insensitive to the new coupling coming from the $a^i$ terms, and the first Friedmann equation (including phenomenological matter) at low energies will be  the usual one \cite{Sotiriou:2009bx,Kiritsis:2009sh} 
with effective gravitational constant as measured in cosmology
\beq
G_{\rm cosmo}=\frac{2}{8\pi (3 \lambda -1)}.
\eeq
The discrepancy between $G_{\rm eff}$ and $G_{\rm cosmo}$ is constrained by measurements of the primordial abundance of He$^4$ \cite{Carroll:2004ai}:
\beq 
 |G_{\rm cosmo}/G_{\rm eff}-1|\lesssim 1/8.
\eeq
Combining this constraint with those in eq.~(\ref{constr}) and the fact that we want $G_{\rm cosmo}>0$, we obtain that there is a triangular region of the $(\eta, \lambda)$ parameter space which is allowed. For $\eta=0$, we obtain the maximum upper bound for $\lambda$ as
\beq
\label{bound}
0<\lambda-1\lesssim  0.1 .
\eeq
Thus, if the theory is to be viable, $\lambda$ will have to be sufficiently close to $1$. 

Continuing the previous analysis to higher orders, we anticipate that the interactions in the perturbation series will scale  as $(|c_\z| M_{\rm pl})^{-n} \sim (\sqrt{|1-\lambda|} M_{\rm pl})^{-n}$ for increasing orders $n$ (as was the case for the previous versions of the theory as well \cite{Blas:2009yd}). Based on the mild bound above, and assuming that no new scale $Z$ is introduced, the strong coupling kicks in at the energy scale $~0.2 M_{\rm pl}$, which is already an order of magnitude lower than the natural cutoff of GR. Note that the experimental constraint used here is not the only one $\lambda$ has to satisfy in principle. In fact, this is just an extremely mild constraint and other constraints coming from Lorentz violations would be much more stringent, requiring that $\lambda$ be even closer to $1$. For example, in \cite{afshordi} the $\eta=0$ version of the theory was studied, yielding a bound $\lambda-1 = 0.003 \pm 0.014$ from the running of the Planck mass. Another bound mentioned in \cite{Blas:2009ck} from the $\alpha_2$ PPN parameter yields  $\lambda-1 \lesssim  10^{-7}$.   These and similar bounds will push the strong coupling scale orders of magnitude lower and definitely deserves further investigation. In any case, the very existence of such of an effective cutoff would imply that this extended theory cannot be considered a UV completion of GR, which is what it was introduced for.  That would be true irrespectively of whether the strong coupling energy scale is low enough to fall within the range in which we can test gravity theories.

Let us summarize. We examined the extended version of Ho\v rava--Lifshitz gravity proposed in \cite{Blas:2009qj} as a way to address viability issues caused by the anomalous dynamics of the scalar degree of freedom that is present in the original version of the theory. Expanding around flat space, it was shown that, even thought the quadratic action of the theory is ``healthy'',  there are cubic interactions that blow up when $\lambda\to 1$. This leaves no space for Lorentz symmetry to be recovered at low energies.  Additionally, even if one is willing to abandon Lorentz symmetry altogether, experimental constraints imply that $\lambda$ has to be close to $1$. The strong coupling scale, which acts as an effective cutoff of the theory, appears to be lower than the cutoff of GR as an effective theory. Considering that the motivation of the model is to constitute a renormalizable theory of gravity,  this casts serious doubts on whether it can be an interesting alternative to GR.

 As mentioned earlier, a possible way out would be to introduce an new scale $Z$ and use this scale instead of the Planck scale for suppressing the higher order operators, so that these operator could take over the perturbative expansion before strong coupling kicks in. However this definitely requires tuning and it would cause naturalness issues, as such a scale would need to be motivated by some additional physics. Note also that the closer to $1$ observations required $\lambda$ to be, the smaller the scale $Z$ should be chosen to be as well. Trying to make $\lambda$ small in order to avoid Lorentz violations related to its present in the kinetic part of the action will effectively lower the scale at which Lorentz violations related to the higher order operators  appear and vice versa.  Thus, in this scenario $\lambda$ will be bound both from above and below, and it remains to be seen how wide is the range of values it can actually take. (As already mentioned, the experimental constraint used here for demonstrative purposes was an extremely mild one.)

Another possible way for the theory to make sense despite the strong coupling could be  through some analogue of the Vainshtein mechanism \cite{Vainshtein:1972sx} in massive gravity.  It is conceivable that such a non-perturbative restoration of the limit  $\lambda\to 1$ will take place in the present theory as well.  Alternatively, one may hope that the renormalization group flow of $\lambda$ will conspire in order for $\lambda$ to be much larger than $1$ around the Planck scale and at the same time   to run very close to $1$ at low energies. If such a tuning can be achieved then the improved theory could constitute an adequate UV completion of GR.

\section*{Acknowledgements}
 The authors would like to thank C.~Charmousis, K.~Koyama and A.~Wang for useful comments and D.~Blas, O.~Pujolas and S.~Sibiryakov for clarifying remarks on the strong coupling scale. TPS was supported by STFC. AP would like to thank DAMTP for their hospitality during the inception of this work.  \\

 \edoc